\documentclass[aps,prc,twocolumn,groupedaddress]{revtex4}
\usepackage{graphicx,epsfig,amsfonts,amssymb,amsbsy}

\usepackage{graphicx}
\usepackage{longtable}
\def\vc#1{\mbox{\boldmath $#1$}}

\begin{document}

%\preprint{}

\title{Density-induced suppression of the $\alpha$-particle condensate in nuclear matter and the structure of $\alpha$ cluster states in nuclei}


\author{Y.~Funaki$^1$, H.~Horiuchi$^2$,  G.~R\"opke$^3$,  P.~Schuck$^{4,5}$, A.~Tohsaki$^2$,  T.~Yamada$^6$}
%\author{ G.~R\"opke$^1$, T.~Yamada$^2$, Y.~Funaki$^3$, P.~Schuck$^4$, H.~Horiuchi$^5$, A.~Tohsaki$^5$}
%\thanks{}
\affiliation{$^1$ Nishina Center for Accelerator-Based Science, The Institute of Physical and Chemical Research (RIKEN), Wako, Saitama 351-0198, Japan}
\affiliation{$^2$ Research Center for Nuclear Physics (RCNP), Osaka University, Osaka 567-0047, Japan}
\affiliation{$^3$ Institut f\"ur Physik, Universit\"at Rostock, D-18051 Rostock, Germany}
\affiliation{$^4$ Institut de Physique Nucl\'eaire, 91406 Orsay Cedex, France}
\affiliation{$^5$ Universit\'e Paris-Sud, F-91406 Orsay-C\'edex, France}
\affiliation{$^6$ Laboratory of Physics, Kanto Gakuin University, Yokohama 236-8501, Japan}
%\affiliation{$^1$ Institut f\"ur Physik, Universit\"at Rostock, D-18051 Rostock, Germany}
%\affiliation{$^2$ Laboratory of Physics, Kanto Gakuin University, Yokohama 236-8501, Japan}
%\affiliation{$^3$ The Institute of Physical and Chemical Research, Wako, Saitama 351-0198, Japan}
%\affiliation{$^4$ Institut de Physique Nucl\'eaire, 91406 Orsay Cedex, France}

%\affiliation{$^5$ Research Center for Nuclear Physics, Osaka University, Osaka 567-0047, Japan}

%\date{\today}

\begin{abstract}
At low densities, with decreasing temperatures, in symmetric nuclear matter $\alpha$-particles are formed,
 which eventually give raise to a quantum condensate with four-nucleon $\alpha$-like correlations (quartetting). 
Starting with a model of  $\alpha$-matter, where undistorted $\alpha$ particles interact via an effective
 interaction such as the Ali-Bodmer potential, the suppression of the 
condensate fraction at zero temperature
 with increasing density is considered. 
Using a Jastrow-Feenberg approach, it is found that the condensate fraction vanishes near saturation density. 
Additionally, the modification of the internal state of the $\alpha$ particle due to medium effects
 will further reduce the condensate.

In finite systems, an enhancement of the $S$ state wave function of the c.o.m. orbital of $\alpha$ particle
 motion is considered as the correspondence to the condensate. 
Wave functions have been constructed for self-conjugate 4$n$ nuclei which describe the condensate state,
 but are fully antisymmetrized on the nucleonic level. 
These condensate-like cluster wave functions have been successfully applied to describe properties
 of low-density states near the $n \alpha$ threshold. 
Comparison with OCM calculations in $^{12}$C and $^{16}$O shows strong enhancement of the occupation
 of the $S$-state c.o.m. orbital of the $\alpha$-particles. 
This enhancement is decreasing if the baryon density increases, similar to the density-induced
 suppression of the condensate fraction in  $\alpha$ matter. 
The ground states of $^{12}$C and $^{16}$O show no enhancement at all, thus a quartetting
 condensate cannot be formed at saturation densities. 

\end{abstract}

%\keywords{}


\maketitle

\section{Introduction}
The properties of nuclear matter at very low densities and low temperatures are dominated by the formation
 of clusters, in particular  $\alpha$ particles. As a well-known concept,  $\alpha$ matter has been
 introduced where symmetric nuclear matter is described by a system of  $\alpha$ particles,
 weakly interacting via effective  $\alpha$ -  $\alpha$ potentials fitted to the scattering phase shifts, 
 such as the Ali-Bodmer interaction potential \cite{Brink,AliBodmer,Wildermut}.

This concept becomes less valid with increasing density. 
First, at finite temperatures other correlations and also single nucleon states appear
 so that we have a mixture of different constituents, described in chemical equilibrium by a mass action law. 
Secondly, at higher densities the internal fermionic structure of the  $\alpha$ particles becomes of
 relevance so that the four-nucleon bound state will be modified by medium effects. A consistent approach
 can be given by  quantum statistical methods \cite{RMS}. 
Using thermodynamic Green functions, the effects of self-energy and Pauli blocking are included
 so that the bound states are dissolved when the density exceeds a critical value. 
For  $\alpha$ particles this critical density, which is also dependent on temperature, is about $\rho_0/5$,
 with $\rho_0 = 0.17 $ fm$^{-3}$ as the saturation density \cite{BSKRS}.

An important phenomenon is the formation of a quantum condensate with strong four nucleon correlations 
 at low temperatures \cite{roepke}. 
At low densities where  $\alpha$ particles are well defined weakly interacting constituents of symmetric nuclear
 matter, we have Bose-Einstein condensation of  $\alpha$ particles. 
With increasing density, quartetting occurs with medium-modified  $\alpha$ particles and disappears
 at a density of about  $\rho_0/3$. Note that quartet condensation has recently also been considered in the context of cold atom physics \cite{quartet}.

The Bose-Einstein condensation for ideal quantum gases is a well-known phenomenon. 
The occupation  of single-particle states is given by the Bose distribution function. 
Below a critical temperature $T_c$, to obey normalization, the state of lowest energy is macroscopically occupied. 
This macroscopically enhanced coherent occupation of the lowest quantum state is denoted as quantum condensate. 
As well known, the fraction of bosons found in the condensate results for the ideal Bose gas 
 as $n_{\rm cond}/n = 1 - (T/T_c)^{3/2}$.

However, this simple picture is no longer valid, if interaction is taken into account. For a 
recent determination of $T_c$ in the interacting case, see Ref.~\cite{SMS}. Here, we want to 
concentrate on interaction effects at zero temperature. 
In general, the condensate fraction is given by the properties of the density matrix which contains a part which factorizes. 
According to Penrose and Onsager \cite{PO}, the quantum condensate in a homogeneous interacting boson system
 at zero temperature is given by the off-diagonal long-range order in the density matrix. 
The non-diagonal density matrix in coordinate representation can be factorized so that
 in the limit $|\vc r - \vc r'| \to \infty$ follows
\begin{equation}
\lim_{|\vc r-\vc r'| \to \infty} \rho(\vc r, \vc r') =  \psi^*_0(\vc r) \psi_0(\vc r') + \gamma(\vc r-\vc r') .
\end{equation}
The last contribution $\gamma(r)$ disappears at large distances, whereas the first contribution determines
 the condensate fraction in infinite matter as
\begin{equation}
n_0 = {\langle \Psi |a^\dagger_0 a_0 | \Psi \rangle \over \langle \Psi | \Psi \rangle}.
\end{equation}
Exploratory calculation of the condensate fraction of $\alpha$ matter will be given in the following Sec.~{II}. 
In contrast to Ref. \cite{roepke,SMS} where the transition temperature $T_c$  for quartetting was considered, 
 we consider here the zero temperature case and analyze the ground state wave function. 
It will be shown that due to the interaction, the condensate fraction is suppressed with increasing density.

An important question is whether such properties of infinite nuclear matter are of relevance for finite nuclei. 
As well known, e.g.,  pairing obtained in nuclear matter within the BCS approach is also clearly seen in finite nuclei.  
Nuclei with densities near the saturation density are well described by the quasiparticle picture
 which leads to the shell model for finite nuclei. 
At low densities, a fully developed $\alpha$ cluster structure similar to $\alpha$ matter is expected. 
Cluster structures in finite nuclei have been well established. 
A density functional approach is able to include correlations and to bridge between infinite matter and finite nuclei.

An interesting aspect of finite nuclei is the enhancement of the occupation of single $\alpha$-particle states similar
 to Bose-Einstein condensation in $\alpha$-particle matter or condensation of bosonic atoms in traps. 
Recently, gas-like states have been investigated in self-conjugate 4$n$ nuclei \cite{thsr}, and a special ansatz
 for the wave function (THSR ansatz), which is similar to the condensate state in infinite matter,
 has been shown to be appropriate in describing low-density isomers. 
In particular, $^8$Be and the Hoyle state of $^{12}$C are well described with this THSR wave function. 
Investigation of states near the four  $\alpha$ threshold in $^{16}$O is in progress \cite{4aTHSR, 4aocm}. 
Predictions for $^{20}$Ne have been given in \cite{nara_tohsaki}.

In Sect.~{III}, we will explain how the suppression of the condensate fraction, calculated
 for infinite nuclear matter, is also seen in the low-density isomers of self-conjugate 4$n$ nuclei,
 in particular for $n=3$ ($^{12}$C).  
First results for $n=4$ ($^{16}$O) are also given. 
General conclusions are drawn in Sec.~{IV}.

\section{Suppression of condensate fraction in $\alpha$ matter at zero temperature}

The theory of Penrose and Onsager \cite{PO} was first applied to a system with hard core repulsion. 
Depending on the filling factor, the suppression of the condensate was calculated. 
In particular, for liquid $^4$He with a filling factor of 28\% at normal conditions,
 the condensate fraction is reduced to $\approx 8\%$ in good agreement with experimental observations.
To give an estimation for  $\alpha$ matter, with an ``excluded volume'' of about 20 fm$^3$ \cite{LS},
 such a filling factor of 28 \% would  arise at $\approx \rho_0/3$ so that a substantial reduction
 of the condensate fraction already below saturation densities is expected for  $\alpha$ matter.

Within a more systematic approach, we follow the work of Clark et al. \cite{Clark}. 
We calculate the reduction of the condensate fraction as function of the baryon density within perturbation theory. 
A uniform Bose gas of $\alpha$ particles, interacting via the potential $V_\alpha(r)$,
 is considered, disregarding any change of the internal structure of the  $\alpha$ particles at increasing density. 
In particular, the dissolution of the  $\alpha$ particle as a four-nucleon bound state
 because of the Pauli blocking is not taken into account. 

The simplest form of a trial wave function incorporating the strong spatial correlations implied by
 the interaction potential is the familiar Jastrow choice,
 $\psi(\vc r_1,\dots, \vc r_A) = \prod_{i < j} f(| \vc r_i - \vc r_j|)$. 
Within our exploratory calculation we consider the lowest approximation with respect to the density
 in order to show the tendency of condensate suppression due to the interaction. 
Normalization gives for the variational function the constraint 
\begin{equation}
\label{norm}
4 \pi \rho_\alpha \int_0^\infty [f^2(r)-1] r^2 dr = -1\, ,
\end{equation}
$\rho_\alpha=\rho/4$ being the density of $\alpha$ particles . 

In the low density limit, the binding energy per $\alpha$-particle is given by 
\begin{equation}
\label{energy}
E[f] = 2 \pi  \rho_\alpha \int_0^\infty \left\{ {\hbar^2 \over 4 M} \left({\partial f(r) \over \partial r } \right)^2 + V_\alpha(r) f^2(r) \right\} r^2 dr,
\end{equation}
$M$ being the nucleon mass. 
The condensate fraction is calculated according to
\begin{equation}
\label{n0}
n_0=\exp \left\{ - 4 \pi \rho_\alpha \int_0^\infty [f(r)-1]^2 r^2 dr \right\}.
\end{equation}
Note that these approximations \cite{Clark} only hold in the low-density limit. 
At higher densities, the pair correlation function has to be evaluated. 
A more advanced approach based on a HNC calculation has been given by Clark, Ristig and others, see \cite{Clark,Ristig}.

For the evaluation of the condensate fraction (\ref{n0}) we use the Ali-Bodmer $\alpha$-$\alpha$
 interaction potential \cite{AliBodmer}
\begin{equation}
\label{potential}
V_\alpha(r) = 457 {\rm e}^{-(0.7 r/{\rm fm})^2} {\rm MeV} -130 {\rm e}^{-(0.475 r/{\rm fm})^2} {\rm MeV} .
\end{equation}
According to Johnson and Clark \cite{Clark} we choose the variational function as
\begin{equation}
\label{function}
f(r)=(1-{\rm e}^{-a r}) (1+ b {\rm e}^{-a r}+c {\rm e}^{-2 a r}).
\end{equation}
After determining the parameters $a,b,c$ from the minimum of energy \cite{RS2006},
 the condensate fraction can be evaluated, see Fig. 1.

%%%%%%%%%%%%%%%%%%%%%%%%%%%%%%%%%%%%%%%%%%%%%%%%%%%%%%%%%%%%%%%%%%%%%%%%%%%%%%%%%%%%%%%%%%%%%%%%%%%%%
\begin{figure}[ht] 
\centerline{\psfig{figure=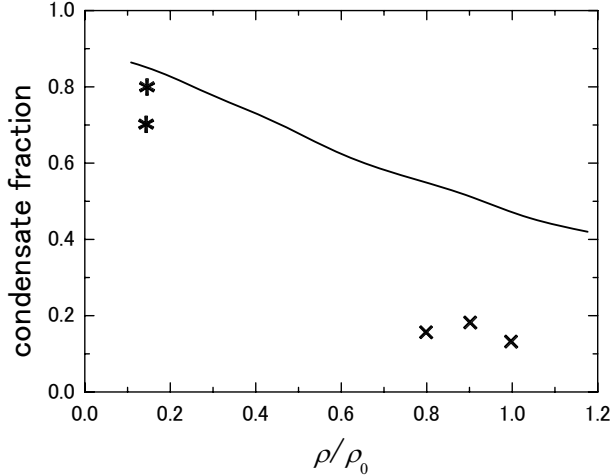,width=10cm,angle=-0 } } 
\caption{
Reduction of condensate fraction in $\alpha$ matter with increasing baryon density
 ($\rho_0$ denotes the saturation density). 
Full line - perturbation approach, crosses - HNC calculations by Johnson and Clark~\cite{Clark},
 stars - Hoyle state (see Sec.~III).
} 
\label{fig1} 
\end{figure}  
%%%%%%%%%%%%%%%%%%%%%%%%%%%%%%%%%%%%%%%%%%%%%%%%%%%%%%%%%%%%%%%%%%%%%%%%%%%%%%%%%%%%%%%%%%%%%%%%%%%%%

In Fig. 1, the full line represents the result for the condensate fraction as function of the baryonic density
 according to the perturbative treatment. 
In the zero density limit this fraction is expected to go to 1. 
Calculations performed by Johnson and Clark \cite{Clark} using a HNC calculation
 for the pair distribution function are given by crosses, showing a stronger suppression of the condensate fraction near the saturation density.

As found from the calculation of the critical temperature for the formation of a quartetting
 condensate \cite{roepke}, we expect that near the saturation density the condensate fraction will disappear. 
For this, we have not only to take into account the HNC type improvement of the pair distribution function,
 but also the Pauli blocking effects which modify the internal structure of the  $\alpha$ particle
 so that the use of the Ali-Bodmer interaction potential is no longer justified. 
Recently, improved versions of the $\alpha$-$\alpha$ interaction have been proposed \cite{Malik}. 
Three-$\alpha$ forces have been considered in order to give a better estimation for the critical point
 of $\alpha$ matter which should be positioned below saturation density \cite{John}. 
Thus, the repulsive part of the $\alpha$-$\alpha$ interaction (which also is a consequence
 of the Pauli blocking with respect to the internal nucleonic structure) is only a part of
 the suppression of the condensate, which is described here. 

Another effect is the medium modification of the internal structure of the $\alpha$ particle
 as well as of the interaction which can be elaborated within a cluster-mean field approximation \cite{RMS}.
The dissolution of $\alpha$-like bound states due to Pauli blocking has been evaluated for
 an uncorrelated medium solving the Faddeev-Yakubowsky equation \cite{BSKRS}. 
It has been shown \cite{roepke} that the four-particle correlations in the condensate disappear
 due to Pauli blocking at around $\rho_0/3$ within a variational approach,
 approximating the 4-nucleon wave function by the solution of the two-particle problem
 and describing the relative c.o.m. motion by a Gaussian wave function. 
Therefore, a medium dependent $\alpha$-$\alpha$ interaction of the Ali-Bodmer type 
 may be expected to account for the features of this effect in an exploratory way. 
In principle, an {\em ab initio} calculation based on interacting nucleons should be performed,
 with Green functions, variational, or AMD techniques.

%%%%%%%%%%%%%%%%%%%%%%%%%%%%%%%%%%%%%%%%%%%%%%%%%%%%%%%%%%%%%%%%%%%%%%%%%%%%%%%%%%%%%%%%%%%%%%%%
\begin{figure}[ht] 
\centerline{\psfig{figure=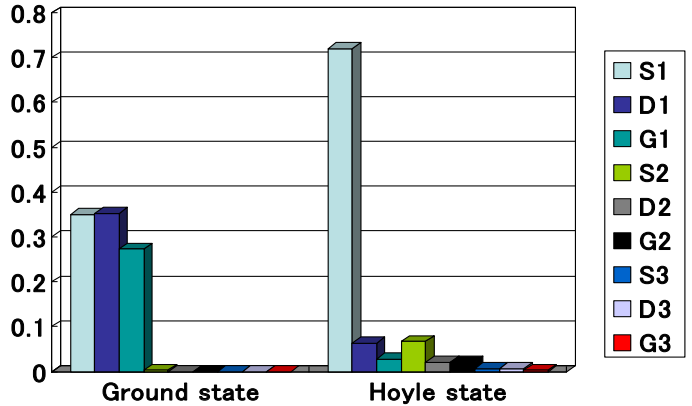,width=10cm,angle=-0 } } 
\caption{
Occupation of the single-$\alpha$ orbitals of the ground state of $^{12}$C compared
 with the Hoyle state~\cite{Yamada}. For explanation see the text.} 
\label{fig2} 
\end{figure}  
%%%%%%%%%%%%%%%%%%%%%%%%%%%%%%%%%%%%%%%%%%%%%%%%%%%%%%%%%%%%%%%%%%%%%%%%%%%%%%%%%%%%%%%%%%%%%%%%

%%%%%%%%%%%%%%%%%%%%%%%%%%%%%%%%%%%%%%%%%%%%%%%%%%%%%%%%%%%%%%%%%%%%%%%%%%%%%%%%%%%%%%%%%%%%%%%%
\begin{figure}[ht] 
\centerline{\psfig{figure=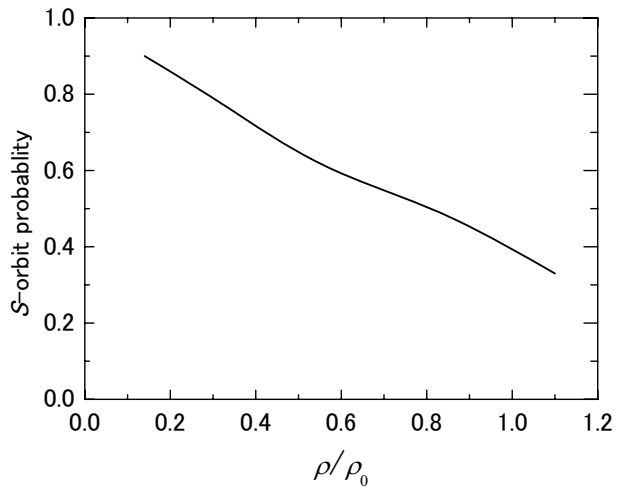,width=10cm,angle=-0 } } 
\caption{Occupation of the $S$1 orbital as function of density using the $3\alpha$ OCM~\cite{Yamada}.} 
\label{fig3} 
\end{figure}  
%%%%%%%%%%%%%%%%%%%%%%%%%%%%%%%%%%%%%%%%%%%%%%%%%%%%%%%%%%%%%%%%%%%%%%%%%%%%%%%%%%%%%%%%%%%%%%%%

\section{Enhancement of cluster c.o.m. $S$ orbital occupation in 4$n$ nuclei}

Signatures akin to Bose-Einstein condensation should arise already in finite nuclei. 
Low-density states of self-conjugate 4$n$ nuclei clearly show an $\alpha$ cluster structure, 
 in particular for $n$ = 2 and $n$ = 3 (Hoyle state). 
The counterpart of a condensate in infinite $\alpha$ matter, where the occupation of the ground state
 is enhanced and becomes of the same order as the total particle number,
 will be the enhancement of the occupation number of a single-$\alpha$ orbital of the $\alpha$-clusters in a low density state of the nucleus. 

The $\alpha$ clustering nature of the nucleus $^{12}$C has been studied by 
 many authors using various approaches\ \cite{carbon}. 
Among these studies, solving the fully microscopic three-body problem of $\alpha$ 
 clusters gives us the most important and reliable theoretical information of 
 $\alpha$ clustering in $^{12}$C within the assumption that no $\alpha$ 
 cluster is distorted or broken except for the change of the size parameter 
 of the $\alpha$ cluster's internal wave function. 
First solutions of the microscopic 3$\alpha$ problem where the antisymmetrization 
 of nucleons is exactly treated, have been given by Uegaki et al.~\cite{uegaki} and
 by Kamimura et al.~\cite{kamimura}.  
In those works, the $^{12}$C levels are described by the wave function of the form 
 ${\cal A} \{\chi({\vc s},{\vc t}) \phi_\alpha^3\}$ with ${\cal A}$ 
 standing for the antisymmetrizer, $\phi_\alpha^3 \equiv \phi(\alpha_1) 
 \phi(\alpha_2) \phi(\alpha_3)$ for the product of the internal wave 
 functions of three $\alpha$ clusters, and ${\vc s}$ and ${\vc t}$ for the 
 Jacobi coordinates of the center-of-mass motion of three $\alpha$ clusters. 
Here $\phi(\alpha_i)$ ($i=1,2,3$) is the internal wave function of the $\alpha$-cluster
 $\alpha_i$ having the form $\phi(\alpha_i) \propto \exp [ -(1/8b^2) \sum_{m>n}^4 ({\vc r}_{i,m} - {\vc r}_{i,n})^2 ]$. 
The wave function $\chi({\vc s},{\vc t})$ of the relative motion of 3 $\alpha$ clusters
 is obtained by solving the energy eigenvalue problem of 
 the full three-body equation of motion; $\langle \phi_\alpha^3 | (H-E) | {\cal A} \{\chi({\vc s},{\vc t}) \phi_\alpha^3\} \rangle = 0$, 
 where $H$ is the microscopic Hamiltonian consisting of the kinetic energy, effective two-nucleon potential,
 and the Coulomb potential between protons. 
The difference between the works by Uegaki et al.~and Kamimura et al.~lies
 in the adopted effective two-nucleon force,
 besides the differing techniques of solution.

Both calculations by Uegaki et al.~and Kamimura et al.~reproduced reasonably 
 well the observed binding energy and r.m.s. radius of the ground $0^+_1$ 
 state which is the state with normal density, while they both predicted 
 a very large r.m.s. radius for the second $0^+_2$ state which is larger 
 than the r.m.s. radius of the ground $0^+_1$ state by about 1 fm, i.e. by over 30\%. 
The observed $0^+_2$ state lies slightly above the 3$\alpha$ breakup 
 threshold and the energies of the calculated $0^+_2$ state reproduced 
 reasonably well the observed value although the value by Uegaki et al. 
 is slightly higher than the 3$\alpha$ breakup threshold by 
 about 1 MeV.  
The second $0^+$ state of $^{12}$C is well known as the key state for the synthesis of $^{12}$C in stars (Hoyle state) 
 and also as one of the typical mysterious $0^+$ states in light nuclei 
 which are very difficult to understand from the point of view of the shell model \cite{nocore}. 

Alternatively, the $0^+_2$ state with dilute density can be described by a gas-like structure
 of $3\alpha$-particles which interact weakly among one another, predominantly in relative $S$ waves.  
The $S$-wave dominancy in the $0^+_2$ state structure had been already suggested 
 by Horiuchi on the basis of the 3$\alpha$ OCM (orthogonality condition model) calculation~\cite{hori}. 
It should be mentioned that not only the binding energy, 
 but also other properties of the  $0^+_2$ state such as electron scattering form factors are well described
 within the calculations given in Refs. \cite{uegaki,kamimura,hori}.

Recently, based on the investigations of the possibility of $\alpha$-particle condensation
 in low-density nuclear matter~\cite{roepke}, the present authors 
 proposed a conjecture that near the $n\alpha$ threshold in self-conjugate 
 $4n$ nuclei there exist excited states of dilute density which are composed 
 of a weekly interacting gas of self-bound $\alpha$ particles and which can be considered 
 as an $n\alpha$ condensed state~\cite{thsr}. 
This conjecture was backed by examining the structure of $^{12}$C and $^{16}$O using a new 
 $\alpha$-cluster wave function of the $\alpha$-cluster condensate type.
The new $\alpha$-cluster wave function, which will be denoted as THSR wave function, actually succeeded
 to place a level of dilute density (about one third of saturation density) in each system 
 of $^{12}$C and $^{16}$O in the vicinity of the 3 respectively 4 $\alpha$ 
 breakup threshold, without using any adjustable parameter.  
In the case of $^{12}$C, this success of the new $\alpha$-cluster wave function may 
 seem rather natural, as we explained above. The microscopic 3$\alpha$ cluster models had predicted that 
 the $0^+_2$ in the vicinity of the 3$\alpha$ breakup threshold has a 
 gas-like structure of $3\alpha$-particles which interact weakly with 
 each other predominantly in relative $S$ waves. Having put forward that Hoyle like states 
in $4n$ self-conjugate nuclei may be a general and common phenomenon is the merit of the work in \cite{thsr}.

The THSR wave function of the $\alpha$-cluster condensate 
 type used in Ref.~\cite{thsr} represents a condensation of $\alpha$-clusters 
 in a spherically symmetric state. 
This is clearly seen by the following expression
\begin{equation}
|\Psi \rangle = {\cal P} {(C^\dagger_\alpha)}^n |{\rm vac}\rangle, \label{eq:cond}
\end{equation}
with
\begin{equation}
\langle1234 |C^\dagger_\alpha  |{\rm vac}\rangle =  \Phi(\vc P) \delta_{\vc P,\vc p_1+\vc p_2+\vc p_3+\vc p_4} 
\phi_\alpha(1234) a_1^\dagger a_2^\dagger a_3^\dagger a_4^\dagger,
\end{equation}
$ \Phi(\vc P)$ describing the c.o.m. motion of the $\alpha$ cluster, and $\phi$ the internal wave function
 of the four-nucleon cluster. 
The operator $\cal{P}$ is projecting out the total c.o.m. motion of the $4n$ nucleus. 
In the limit of infinite nuclear matter, the $\Phi$ orbitals are plane waves, and the projection operator
 $\cal{P}$ can be neglected. In the case considered here, the use of Gaussians allows the explicit separation
 of the c.o.m. motion of the four-nucleon cluster as well as of the whole $4n$ nucleus. It should also be noted 
that Eq. (\ref{eq:cond}) contains two limits exactly: the one of a pure Slater determinant relevant 
at higher densities and the one of a $100$ percent ideal $\alpha$-particle condensate in the dilute 
limit \cite{thsr}. All intermediate scenarios are also correctly covered.

The present authors extended the wave 
 function so that it can describe the $\alpha$-cluster condensate with spatial 
 deformation \cite{fhtsr}. 
They applied this new wave function to $^8$Be and succeeded to reproduce not only the binding energy of the ground 
 state but also the energy of the excited $2^+$ state. 
In addition, they found that although the effect of the spatial deformation is not large, 
 the introduction of the spatial deformation brought forth a 100 \% overlap of the THSR wave function
 with the ``exact'' wave function given by the microscopic 2$\alpha$ cluster model which solves 
 the 2$\alpha$-cluster equation of motion, 
 $\langle \phi_\alpha^2 | ( H - E ) | {\cal A} \{ \chi ( {\vc r} ) \phi_\alpha^2 \} \rangle = 0$.  
This fact forces us to modify our understanding of the $^8$Be structure from the 2$\alpha$ ``dumb-bell''
 structure to the 2$\alpha$ dilute (gas-like) structure.
It was shown that the $0^+_2$ wave function of $^{12}$C which was obtained long time ago
 by solving the full three-body problem of the microscopic $3\alpha$ cluster model is almost completely equivalent to 
 the wave function of the $3\alpha$ THSR state.  
This result gives us strong support to our opinion that the 
 $0^+_2$ state of $^{12}$C has a gas-like structure of $3\alpha$ clusters with ``Bose-condensation''. 
The rms radius for this THSR state was calculated as ${R(0^+_2)}_{\rm THSR}$ = 4.3 fm which fits well with 
experimental data for the form factor of the Hoyle state, 
see Ref. \cite{Neff}. It confirms the assumption of low density as a prerequisite for the formation of 
an $\alpha$-cluster structure for which the Bose-like enhancement of the occupation of the $S$ orbit is possible.

Recently, a fermionic AMD calculation based on nucleons with effective interactions has been
 performed \cite{Neff} which supports the applicability of the THSR state to describe
 the Hoyle state. 
It is found hat the form factor calculated for the $0^+_2$ state of $^{12}$C coincides
 with the form factor obtained from the THSR wave function. 
In particular, the low density of nucleons, the formation of four-nucleon clusters and the dominant
 contribution of the gas-like distribution has been confirmed.

A very interesting analysis of the applicability of the THSR wave function
 can be performed by comparing with stochastic variational calculations \cite{Suzuki}
 and OCM calculations \cite{Yamada}. 
The $\alpha$ density matrix $\rho(\vc{r},\vc{r}')$ defined by integrating out of the total
 density matrix all intrinsic $\alpha$-particle coordinates, is diagonalized to study
 the single-$\alpha$ orbits and occupation probabilities in $^{12}$C states.
Fig. 2 shows the occupation probabilities of the $L$-orbits with $S$, $D$ and $G$ waves 
 belonging to the $k$-th largest occupation number
 (denoted by $Lk$), for the ground and Hoyle state of $^{12}$C
 obtained by diagonalizing the density matrix $\rho(\vc{r},\vc{r}')$.
We found that in the Hoyle state the $\alpha$-particle $S$ orbit with zero node ($S1$ in Fig.~2)
 is occupied to more than 70~\% by the three $\alpha$-particles (see also Ref.~\cite{Suzuki} and Fig.~1).
Taking into account the finite size of the nucleus, a reduction of the condensate fraction from 100 \%
 to about 70 \% is not surprising, and the remaining fraction (about 30 \%) is due to higher orbits
 originating from antisymmetrization among nucleons.
This huge percentage means that an almost ideal $\alpha$-particle condensate is realized
 in the Hoyle state. 
One should remember that superfluid $^4$He has only 8 percent of the particles in the condensate, 
what represents a macroscopic amount of particles nonetheless. Please also note that the $S$-wave 
occupancy of the Hoyle state is at least by a factor ten larger than the occupancy of any other 
state (Fig. 2). Independent of the absolute occupancy of the $S$-wave state, this is a clear 
signature of quantum coherence, i.e. of condensation.
  
On the other hand, in the ground state of $^{12}$C, the $\alpha$-particle occupations
 are equally shared among $S1$, $D1$ and $G1$ orbits, where they have two, one, and zero nodes,
 respectively, reflecting the SU(3)$(\lambda\mu)=(04)$ character of the ground state \cite{Yamada}.
This fact thus invalidates a condensate picture for the ground state.  
 
To get a more extended analysis, OCM calculations have been performed \cite{Yamada} for studying the density
 dependence of the $S$-orbit occupancy in the $0^+$ state of $^{12}$C on the different densities
 $\rho/\rho_0 \sim (R{(0^+_1)}_{\rm exp}/R)^3$, in which the rms radius ($R$) of $^{12}$C is taken 
as a parameter and $R{(0^+_1)}_{\rm exp} $=2.56 fm.
A Pauli-principle respected OCM basis $\Psi^{\rm OCM}_{0^+}(\nu)$ with a size parameter
 $\nu$ is used, in which the value of $\nu$ is chosen to reproduce a given rms radius $R$ of $^{12}$C, 
 and the $\alpha$ density matrix $\rho(\vc{r},\vc{r}')$ with respect to $\Psi^{\rm OCM}_{0^+}(\nu)$
 is diagonalized to obtain the $S$-orbit occupancy in the $0^+$ wave function.      
The results are shown in Fig. 3.
The $S$-orbit occupancy is $70\sim 80$~\% around $\rho/\rho_0\sim (R{(0^+_1)}_{\rm exp}/R{(0^+_2)}_{\rm THSR})^3 = 0.21$, while
 it decreases with increasing $\rho/\rho_0$ and amounts to about $30\sim40$ \% in the saturation
 density region.
A smooth transition of the $S$-orbit is observed from the zero-node $S$-wave nature ($\rho/\rho_0\simeq0.2$) to a
 two-node $S$-wave one ($\rho/\rho_0\sim1$) with increasing $\rho/\rho_0$~\cite{Yamada}.  
The feature of the decrease of the enhanced occupation of the $S$ orbit is in striking correspondence
 with the density dependence of the condensate fraction calculated for nuclear matter (see Fig. 1). 

An interesting item is whether there exist other nuclei showing the Bose
condensate-like enhancement of the $S$-orbit occupation number. Then, the
suppression of the condensate with increasing density is also of relevance
for those nuclei. After we discussed the case of $^{12}$C corresponding to
$n=3$ we will now shortly
discuss the situation in the next nucleus $^{16}$O corresponding to
$n=4$, where great efforts are performed recently to investigate
low-density excitations in the $0^+$ spectrum in theory as well as in
experiments.

% In Ref. \cite{thsr}, THSR states for $^{16}$O have been considered, and
% a spectrum of $0^+$ states has been observed. The second excited state
% ${(0^+_3)}_{\rm THSR}$ was considered as candidate for a Bose-condensate 
% like state, with energy near the 4 $\alpha$ threshold and a rms radius 
%  $R{(0^+_3)}_{\rm THSR} = 3.97$ fm. Compared with the ground state of $^{16}$O 
% with the experimental rms radius $R{(0^+_1)}_{\rm exp} = 2.73$ fm, we have
% an expanded state where the relative density $\rho/\rho_0$ is estimated by
% $(R{(0^+_3)}_{\rm THSR}/R{(0^+_1)}_{\rm exp})^3 = 0.33$. This value is larger
% than the corresponding value 0.2 for the Hoyle state, and therefore a condensate 
% is stronger suppressed at this density. Thus we conclude now that this THSR state
% cannot be realized in a sufficient approximation by a real excitation of  $^{16}$O,
% because in addition to the $S$ orbit also other orbits will significantly
% contribute to the 4 $\alpha$ wave function.

In analogy to the aforementioned OCM calculation for $^{12}$C \cite{Yamada}, we recently 
performed a quite complete OCM calculation also for $^{16}$O, including many of 
the cluster configurations, just mentioned (a full account will be 
given in a separate publication \cite{4aocm}). We were able to reproduce the full spectrum 
of $0^+$ states with $0_2^+$ at $6.4$ MeV, $0_3^+$ at $9.4$ MeV, $0_4^+$ at $12.6$ MeV, 
$0_5^+$ at $14.1$ MeV, and $0_6^+$ 
at $16.5$ MeV. Also the rms radii are obtained. The largest values are found as
 $R{(0^+_6)}_{\rm OCM}$ = 5.6 fm, followed by  $R{(0^+_4)}_{\rm OCM}$ = 4.0 fm.
We tentatively make a one to one 
correspondence of those states with the six lowest $0^+$ states of the 
experimental spectrum. In view of the complexity of the situation, the 
agreement can be considered as very 
satisfactory. The analysis of 
the diagonalization of the $\alpha$-particle density matrix $\rho({\vc 
r},{\vc r}')$ (as was done in Ref.~\cite{Yamada}) showed that the newly discovered $0^+$ 
state at $13.6$ MeV \cite{Wakasa}, as well as the well known $0^+$ state at $14.01$ MeV, 
corresponding to our states at $12.6$ MeV and $14.1$ MeV, respectively, have, 
contrary to what we assumed previously \cite{nupecc}, very 
little condensate occupancy of the zero-node $S$-orbit (about $20$ percent). On the 
other hand, the sixth $0^+$ state at $16.5$ MeV calculated energy, to be 
identified with the experimental state at $15.1$ MeV, has $61$ percent of the 
$\alpha$ particles being in the zero-node $S$-orbit. 

These results confirm our statement that the $\alpha$-particle condensate in nuclear matter is suppressed
with increasing density and, consequently, a well developed condensate state in nuclei 
can be expected only at very low densities. For $^{16}$O, the relative  densities $\rho/\rho_0$ are estimated as
$(R{(0^+_1)}_{\rm exp}/R{(0^+_4)}_{\rm OCM})^3 = 0.32$ and  $(R{(0^+_1)}_{\rm exp}/R{(0^+_6)}_{\rm OCM})^3 = 0.12$.
Therefore we expect a significant enhancement of the $S$ orbit occupation number only for the  $0_6^+$ state, 
in full agreement with the OCM calculation cited above.
The very large radius of that state is again a clear indication of an $\alpha$-particle gas (Hoyle)-like state, 
and the THSR wave function is expected to describe this state in a sufficient approximation. 
%Work in determining the 
%next THSR state ${(0^+_4)}_{\rm THSR}$ which will be a good candidate for a Bose-condensate like state in nuclei is in progress.
Work in determining the complete spectrum of THSR states in  $^{16}$O showing the relevance of a Bose-condensate 
like state is in progress \cite{4aTHSR}.

\section{Conclusions}

Multiple successful theoretical investigations, concerning the Hoyle 
state in $^{12}$C, have established, beyond any doubt, that it is a dilute 
gas-like state of three $\alpha$-particles, held together only by the 
Coulomb barrier, and describable to first approximation by a wave function 
of the form $(C_{\alpha}^\dagger)^3|{\rm vac}\rangle$ where the three bosons 
$(C_{\alpha}^\dagger)$ are condensed into the $S$-orbital. There is no objective 
reason, why in $^{16}$O,$^{20}$Ne,$\cdots$ there should not exist similar 'Hoyle'-like 
states. At least the calculations with THSR and OCM approaches show this 
to be the case, systematically. 
%In this work, we give preliminary results 
%of a complete OCM calculation which reproduces the six first $0^+$ states of 
%$^{16}$O to rather good accuracy. In 
%that calculation the $0_6^+$ state at $16.5$ MeV, to be identified with the 
%experimental $0^+$ state at $15.1$ MeV, shows the characteristics typical for 
%a Hoyle-like state, that is high $\alpha$-particle $0S$-wave occupancy combined with an unusually large radius. 
In this work, we give preliminary results of a complete OCM calculation which reproduces
 the six first $0^+$ states of $^{16}$O to rather good accuracy. 
In that calculation the $0_6^+$ state at $16.5$ MeV, which might be identified with the experimental $0^+$
 state at $15.1$ MeV, shows the characteristics typical for a Hoyle-like state, that is
 high $\alpha$-particle $S$-wave occupancy combined with an unusually large radius.
 
Therefore, the main quantity for the formation of an $\alpha$ cluster state is the density
 which should be low. 
Then, the occurrence of a THSR state where all $\alpha$ particles occupy the same orbit with respect
 to the c.o.m. motion is an interesting effect which corresponds to the formation of an $\alpha$-particle
 condensate in symmetric nuclear matter. 
The condensate fraction is decreasing with increasing density because of correlations, as known from
 interacting Bose systems. 
In addition, the internal structure of the  four nucleon cluster is changed due to Pauli
 blocking if density is increasing.

Only in the very low density limit, the  $\alpha$ particles may be considered as independent
 bosons moving relatively free like quasi particles. 
A mean field approach of the interaction which is assumed to be week would give
 a Gross-Pitaevskii equation \cite{YS}. 
Then we can apply the approach of a non-interacting Bose gas where the $\alpha$ particles
 may occupy the same c.o.m. orbital. The enhanced occupation of the ground state (plane wave)
 in infinite matter is the standard description of Bose-Einstein condensation. 
This corresponds, in finite nuclei, to the enhanced occupation of the same orbital for
 the c.o.m. motion so that the THSR state will be a good approximation for the many-nucleon
 wave function. 
We stress the similarity to two-particle pairing where the concept of a BCS state was successfully
 applied to finite nuclei. The question of finite number of Cooper pairs in the nuclear BCS 
state is also to be considered in analogy with the finite number of $\alpha$-particles in the THSR state.

With increasing contribution of the interaction, e.g. with increasing density, the condensate state
 becomes more complex. 
Calculations in infinite matter ($T=0$) show that the condensate state becomes 
increasingly non-ideal (the condensate fraction is smaller than one).
The same is also observed in OCM calculations for finite nuclei where with increasing density
 the condensate state becomes gradually depleted. 
We conclude that there are similarities between the structure of the ground state wave function of 
 $\alpha$ matter and the  $\alpha$ gas like states in finite nuclei.

In addition to the effect of interaction, mixing higher states of c.o.m. orbits to the ground
 state wave function, there is also the dissolution of the internal wave function of the  $\alpha$ 
 particle due to medium effects. 
The transition from the cluster picture with well defined  $\alpha$ states to a shell model
 where nucleons move independently in a mean field is also reproduced in harmonic oscillator
 approximation, but needs a first principle approach to calculate
 the many-nucleon wave function.

These results are also of relevance for other phenomena which arise if the local density approach
 is used. 
Low density matter arises in the halo of heavy nuclei so that preformation of $\alpha$-clusters
 is an interesting issue there, but also in heavy ion reactions or during supernova explosions. 
Cluster condensation very likely will soon also become an important subject in cold atom physics. 
Theoretical investigations already have appeared \cite{quartet}. 
So far nuclear physics is at the forefront of this subject. 

\section*{Acknowledgements}
The present authors (G.R. and T.Y.) express thanks to M. L. Ristig and J. W. Clark, and also to the organizers
 of the International Workshop on Condensed Matter Theories (CMT31) at Bangkok, Thailand, 
 (V. Sa-yakanit: chairperson), where this paper was completed.

%\bibitem{mori}
% H. Morinaga, Phys. Rev. {\bf 101}, 254 (1956); Phys. Lett. {\bf 21}, 
% 78 (1966).
%\bibitem{brink}
% D. M. Brink, {\it Proc. Int. School Phys. Enrico Fermi} {\bf 36} 
% (Academic Press, New York and London, 1966), p.247.
%\bibitem{volkov}
% A. B. Volkov, Nucl. Phys. {\bf 74}, 33 (1965).

%@ARTICLE{feyn54,
%   author = "R. P. Feynman",
%   year = "1954",
%   journal = "Phys.\ Rev.",
%   volume = "94",
%   pages = "262"
%}

%@ARTICLE{epr,
%   author = "A. Einstein and B. Podolsky and N. Rosen",
%   journal = "Phys.\ Rev.",
%   volume = "47",
%   pages = "777"
%}

%@MISC{witten2001,
%   author = "Edward Witten",
%   eprint = "hep-th/0106109"
%}

\end{document}